# Flexible metasurface holograms

**Authors**

J. Burch,[1]* D. Wen,[2] X. Chen,[2] A. Falco[1]

**Affiliations**

[1]SUPA, School of Physics and Astronomy, University of St Andrews, North Haugh, St Andrews, KY16 9SS, UK
[2]SUPA, Institute of Photonics and Quantum Sciences, School of Engineering and Physical Sciences, Heriot-Watt University, Edinburgh, EH14 4AS, UK

**Abstract**

Metasurface holograms are typically fabricated on rigid substrates. Here we experimentally demonstrate broadband, flexible, conformable, helicity multiplexed metasurface holograms operating in the visible range, offering increased potential for real life out-of-the-lab applications. Two symmetrically distributed holographic images are obtained when circularly polarized light impinges on the reflective-type metasurface positioned on non-planar targets. The two off-axis images with high fidelity are interchangeable by controlling the helicity of incident light. Our metasurface features the arrangement of spatially varying gold nanorods on a flexible, conformable epoxy resist membrane to realize a Pancharatnam-Berry phase profile. These results pave the way to practical applications including polarization manipulation, beam steering, novel lenses, and holographic displays.

# MAIN TEXT

## Introduction

Metasurfaces (MSs) are ultra-thin artificial materials made of individual structures, called meta-atoms. These meta-atoms dictate the optical properties of the resulting metamaterial with their specific shape, size, orientation, and arrangement. MSs are essentially planar and thus much simpler to fabricate than bulk MMs [1], with significant consequences for practical applications.

One of the prominent features of MSs is that they permit to control the phase and amplitude of impinging light over scales much smaller than its wavelength [1] [2]. Thus, it is possible to design MSs with extremely complex behavior. Some recent examples include the generalization of Snell's law [3], flat lensing [2, 4], ultra-broadband coherent perfect absorption [5], beam steering [6, 7], ultrathin vortex wave plates for applications in optical tweezers and optical communication systems [8], and wide angle filters [9]. Additionally, the optical response of MSs can be tuned, by varying geometrical or physical parameters, e.g. for to application in polarization manipulation [10], tunable absorption [11, 12], laser steering [6, 7], signal modulation [13], nonreciprocal EIT [14], and time-varying MS and Lorentz nonreciprocity [15].

The capability to engineer the phase of an optical beam with high accuracy and spatial resolution is perfectly suited for holographic applications. This has been exploited to demonstrate several applications in security [16], displays [17] and the storage and manipulation of information [18, 19]. MSs allow for pixel by pixel tailoring of the phase profile of the hologram, allowing for highly efficient designs compared to amplitude only holograms [20, 21]. Holographic MSs use the 2D arrangement of meta-atoms to produce an image from the scattered incident light. Such an image can be computer generated by means of iterative phase reconstruction algorithms, such as Gerchberg-Saxton [22]. Furthermore, these images can embed and exploit novel degrees of freedom, such as the helicity [16], or polarization of the incident light [23, 24], with profound implications e.g. for the analysis of DNA structure [25] and stereochemistry [26].

Realizing holographic metasurfaces on flexible substrates brings about several advantages. Firstly, flexible MSs are compatible with roll-to-roll production and printed electronics [27]. Roll-to-roll production is of great interest for bringing MSs inexpensively, and in high volume, to market, given its relative technological maturity [28-30].

Additionally, flexible MS are conformable to non-planar surfaces, which allows for novel and diverse applications aimed at providing traditional, and dull materials with an advanced photonic layer, e.g. to include ultrathin devices on, clothing, optical fibers [31] [32], contact lenses [33], packaging and bio-inspired applications [34].

Lastly, flexible MSs can also be mechanical tuned through stretching or vibration [1, 35] [36], e.g. to vary the focal length of a metasurface lens [37].

In this paper we designed, fabricated and characterized flexible holographic MSs to project a holographic image. Our holographic images were characterized before and after the lift-off process from a substrate, and placed on a non-perfectly planar substrate. This allowed us to quantify the difference in MS quality in the rigid and flexible regimes. Our holographic MS works in reflection for incident light, and two image designs were tested with helical polarizations.

## Results

### Hologram design

One of the most successful schemes to realize a broadband, broad angle, holographic MS is the reflective geometry as shown in fig. 1. These devices typically consist of three layers, a nano-patterned surface containing the meta-atoms, a spacing layer, and a reflective backplane. The three-layer design acts as a Fabry-Perot cavity to boost phase conversion efficiency, and reduce the dispersion typical of a single MS layer [16, 21]. Using this approach, researchers have achieved efficiencies of 60-80% [16, 21].

The meta-atoms are metallic nanorods, with plasmonic resonances characterized by a fast and slow axis, due to the form factor. By tailoring the exact dimensions of the nanorod to obtain a phase change of $\pi$ between these axes, the nanorod emulates a reflective type half waveplate [21]. In transmission the $\pi$ phase change produces circularly polarized light of the opposite handedness to the incident light, with a phase change of $2\varphi$. Such a half waveplate can thus be used to accurately control the phase of reflected light over a broad wavelength range.

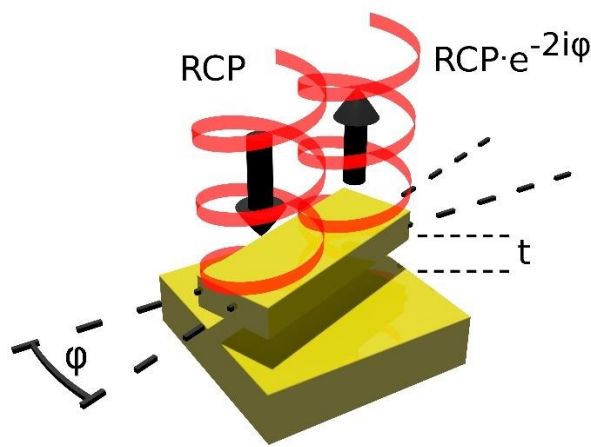

**Fig. 1.** A diagram of the unit cell structure of a nanorod with incident right hand circularly polarized light. The spacing between the nanorod and backplane is denoted as t.

We used the Gerchberg-Saxton algorithm to iteratively retrieve the phase profile to generate the required far field image. Helicity multiplexed holograms differ from conventional linearly polarized holograms in that the two centrosymmetric images can be exchanged based on the helical direction of incident light. This exchange occurs because the sign of the intensity distribution in the far field is dependent on the sign of the phase function of the incident light. An off-axis holographic design was chosen to separate the images from the zeroth order beam. To iteratively retrieve the phase, we assumed a uniform planar source wave, and that there exists a Fourier transform relation between the hologram plane, and the far field. We then encode the phase profile in the hologram plane onto the MS by the angle of gold nanorods point by point.

The retrieved phase profile is realized by 1000×1000 nanorods with a neighboring distance of 300 nm, and an area of 0.3 mm×0.3 mm. To reduce the size of diffraction spots and improve image quality in the far field we constructed a 2×2 periodic array of our retrieved phase profile. As such our MS had a total area of 0.6 mm×0.6 mm. The nanorod spacing was chosen to create an image with a projection range of 95° at 650 nm, as determined by: $d_x = M \Delta P = m\lambda / \tan(a_x / 2)$ and $d_y = N \Delta P = m\lambda / \tan(a_y / 2)$ where $M$ and $N$ are the number of pixels in the hologram for the x and y directions respectively, $\Delta P$ is the pixel size, $m$ and $n$ are the pixels in the projected holographic image, $\lambda$ is the wavelength, $a_x$ and $a_y$ are the angular ranges in x and y respectively. Additionally, the spacing between the nanorods must be small enough to satisfy the Nyquist-Shannon sampling theorem for a continuous phase profile, whilst keeping the distances large enough to minimize cross-talk between the nanorods. To further limit cross-talk between nanorods we used 16 discrete nanorod angles. These angles were chosen to maximize the minimum distance between nanorod corners, whilst keeping frequent sampling across all angles. Each nanorod angle encodes twice the retardation in the Pancharatnam-Berry phase angle of light transmitted through the nanorod.

**MS fabrication**

To fabricate the flexible MS hologram we followed the procedure sketched in fig. 2, and outlined in detail in the materials and methods section.

Firstly we spin coated, on a silicon carrier, the lift off layer. Secondly we spin coated a thick polymer layer to be used as the MS substrate once lift-off has occurred. We then evaporated a gold film onto the sample to act as a reflective backplane. Next we spin coated a thin polymer layer to be used as a spacer between the gold backplane and the MS. We then deposited a thin layer of gold on the sample and spun photoresist on top. After a standard electron beam lithography process, we used a reactive back etch to define gold nanorods on top of our spacing layer. The lift-off layer could then be dissolved to leave a free-floating membrane.

After lift-off we floated the sample on deionized water, where it could be manipulated into its final location. fig. 2E displays an image of a typical region of the MS after lift-off taken by a scanning electron microscope, which shows that the quality of the nanopattern is not affected by the lift-off procedure.

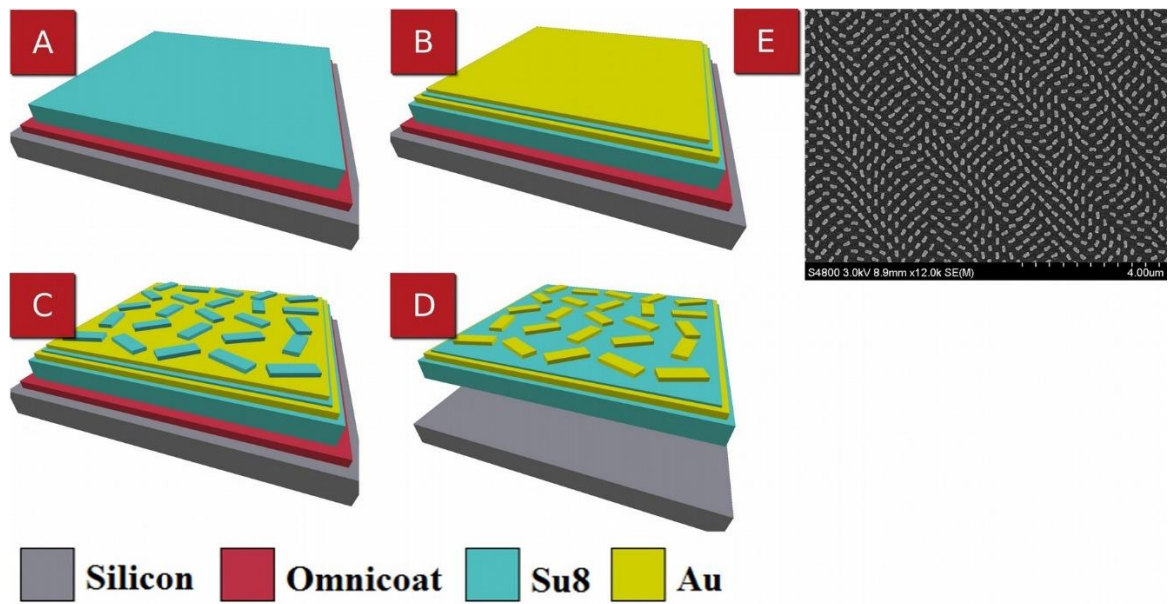

**Fig. 2.** Schematic of the fabrication process and the fabricated flexible MS hologram. **(A)** A silicon substrate is initially coated with an Omnicoat layer and a thick SU8 film by spin-coating. **(B)** A gold film is evaporated onto the SU8 film as a reflective backplane. After that, a spacing layer of SU8 is then spun onto the gold film. Finally, a second gold film is deposited on the SU8 film. **(C)** A resist layer of SU8 is spun onto the final gold layer for the standard electron beam lithography process. Nanorods are defined after the development process and then used as the etching mask. **(D)** Gold nanorods on the top of the sample are then obtained after a reactive ion etch. The Omnicoat layer is then dissolved to leave a free-floating flexible hologram. **(E)** An SEM image of a typical area of the nanorod MS taken after lift-off. It was observed that there were no discernible visual differences between the SEM images of the sample before, and after lift-off (see supplementary fig. S1).

**Experimental setup**

To quantitatively assess the viability of the process, we characterized the efficiency of the holographic MS before, and after lift-off from the rigid silicon substrate. The MS was excited with a SuperK EXTREME supercontinuum laser in the range 570-850 nm, with the setup sketched in fig. 3A and 3B. By passing the laser through a linear polarizer, and a quarter waveplate, we created light with polarizations from linear to circular with the required helicity.

To calculate the optical efficiency of our device we collected the reflected power on one side of the hologram, focusing the image with a lens of f = 25.4 mm on a Thorlabs S130C Photodiode Power Sensor, as shown in fig. 3A. This value was normalized to the incident power, measured by placing the lens and detector directly in front of the quarter waveplate. Since the beam size was larger than the patterned area, we scaled the efficiency by the ratio of the power incident on the hologram and the total incident power. The beam size as function of wavelength was measured using a Thorlabs BC106N-VIS/M CCD beam profiler and the correction factor is shown in supplementary fig. S2. The total efficiency of the hologram was calculated considering the combined two sides of the holographic image. This was done by doubling the efficiency from one side of the holographic image. The efficiency vs wavelength measurements were made in the range 570-850 nm at normal incidence. Efficiency vs angle measurements were made at 650 nm.

For photographing the holographic images, we projected the light incident on the MS onto a paper screen 100 mm from the MS as in fig. 3B. We then photographed the screen with a standard camera. The hole in the screen had a diameter of 3 mm.

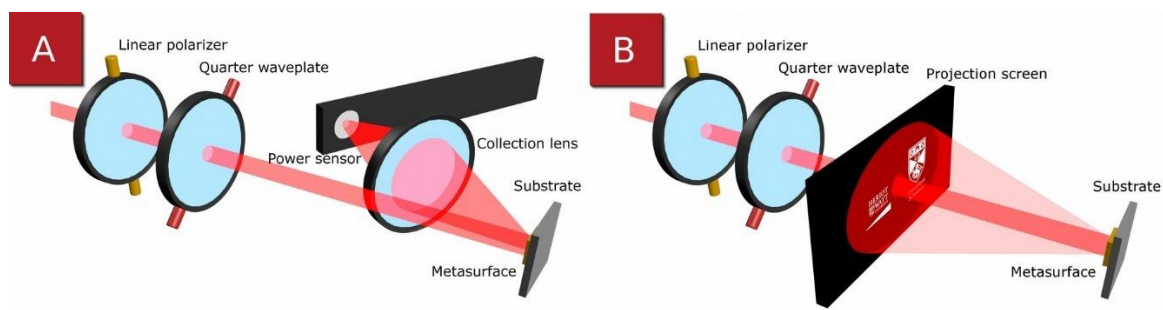

**Fig. 3.** Schematics of the optical set-ups. The helicity of incident light on the MS is controlled by the relative angle between the linear polarizer and the quarter waveplate. In these images right handed circularly polarized light is shown. **(A)** To make efficiency measurements the light scattered by the MS is collected by a f = 25.4 mm lens and the intensity is measured with a Thorlabs S130C Photodiode Power Sensor **(B)** To photograph the holographic image the light scattered from the MS is projected onto a screen.

**Helicity multiplexed hologram**

Fig. 4 shows that our flexible device can form high fidelity helicity multiplexed holograms. Fig. 4A and 4B show the target image for right and left circular polarization respectively. The picture was taken using the setup of fig. 3B. The distortion due to the spherical far field being projected onto a flat plane has been pre-compensated for as detailed in supplementary fig. S3. The cross in the center of fig. 4C and 4D is a simple measurements artefact, due to the MS acting as a square aperture. The movie in the SI shows the dynamic transition between the two polarization states. Panels E-G) of fig. 4 show a zoomed in view of the designed image, and of the images obtained before and after lift-off, respectively. These results demonstrate that the process of lift off does not lead to any visible degradation in the image fidelity and signal to noise ratio.

In fig. 5A, 5B, and 5C we measured the performance of the holographic MS in the visible range from green to red. In the far field the focus of our holographic images is robust to the exact distance to the screen, and their size scales linearly with this distance without distortion. The efficiency of our design decreased roughly linearly with increasing angle as can be seen in figs. 5D and 5E. Angular robustness for small changes is of particular interest for MS tunability. In fig. 5E we show that for right hand circularly polarized light, for a 99% dark image, our device peaks in efficiency around 730 nm at 40% before lift-off, and 37% after lift-off. As such the lift-off process does not significantly impact holographic efficiency. At wavelengths far from 730 nm the dimensions of the nanorods, and the thickness of the spacing layer, are not optimized to produce the $\pi$ phase change and so this efficiency decreases. Furthermore, at shorter wavelengths the interband transitions in gold cause excessive absorption. This reduces the efficiency of the device to the point that holographic images cannot be seen much below 570 nm. Despite this limitation we decided to use gold as the plasmonic material, because unlike silver, it can be back-etched with a reactive ion etch. This was important as RIE defines nanoscale features better than a lift-off procedure on our SU8 polymer spacing layer. For applications requiring the whole visible range gold could be interchanged with silver [16], or aluminum [38], if chlorine is used during the etch. Furthermore, gold is strongly inert which leads to better stability of nano-sized gold features than silver or aluminum.

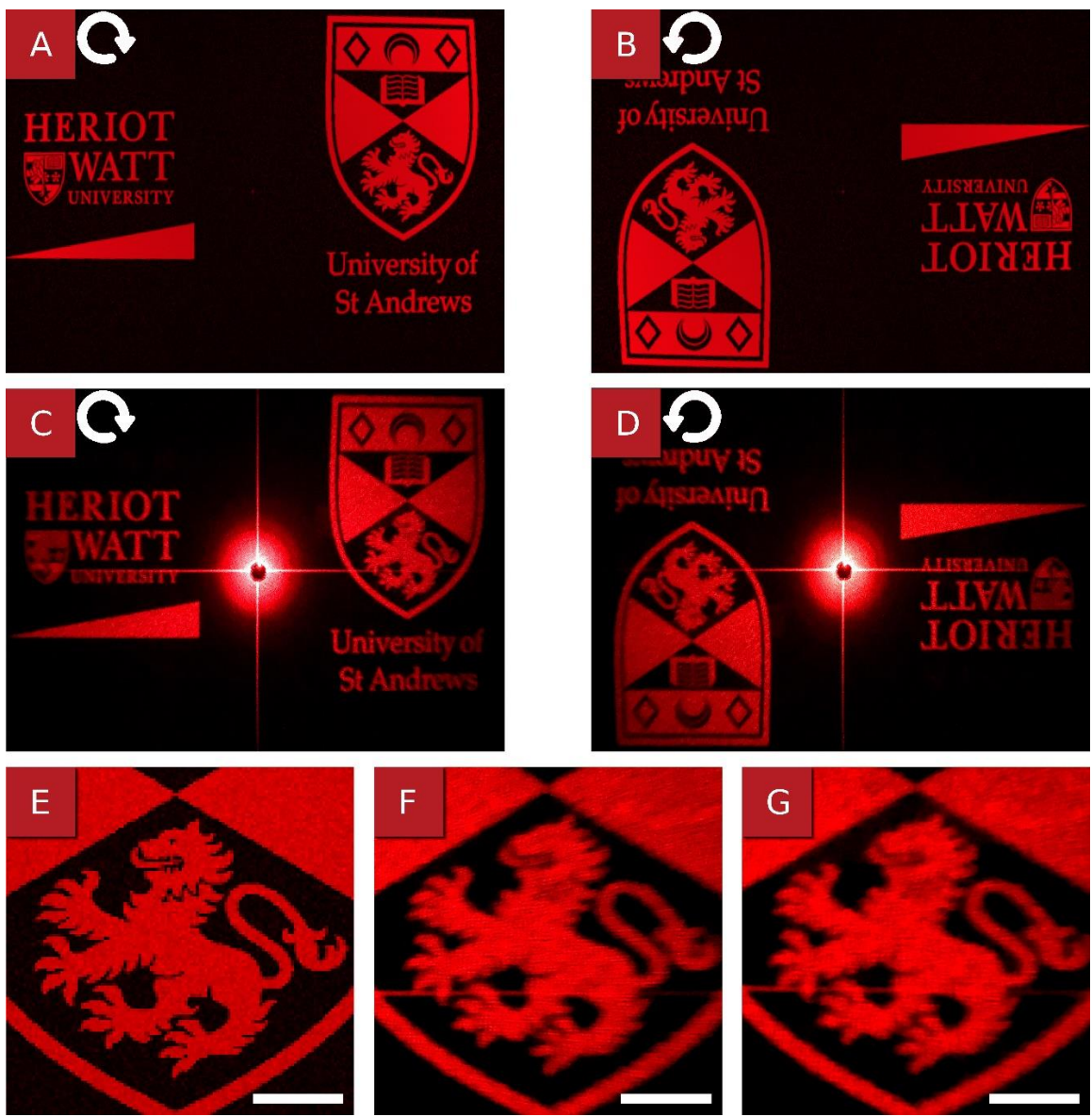

**Fig. 4.** Reconstructed images for the incident light with a wavelength of 650 nm at normal incidence. The simulation results for the incident light with **(A)** right circular polarization and **(B)** left circular polarization and the corresponding experimental results after lift-off shown in **(C)** and **(D)**. **(E)** Simulated close up of the lion. The comparison of experimentally measured close up of the lion **(F)** before and **(G)** after lift-off from the substrate. The scale bar is 10 mm with a MS to screen spacing of 100 mm.

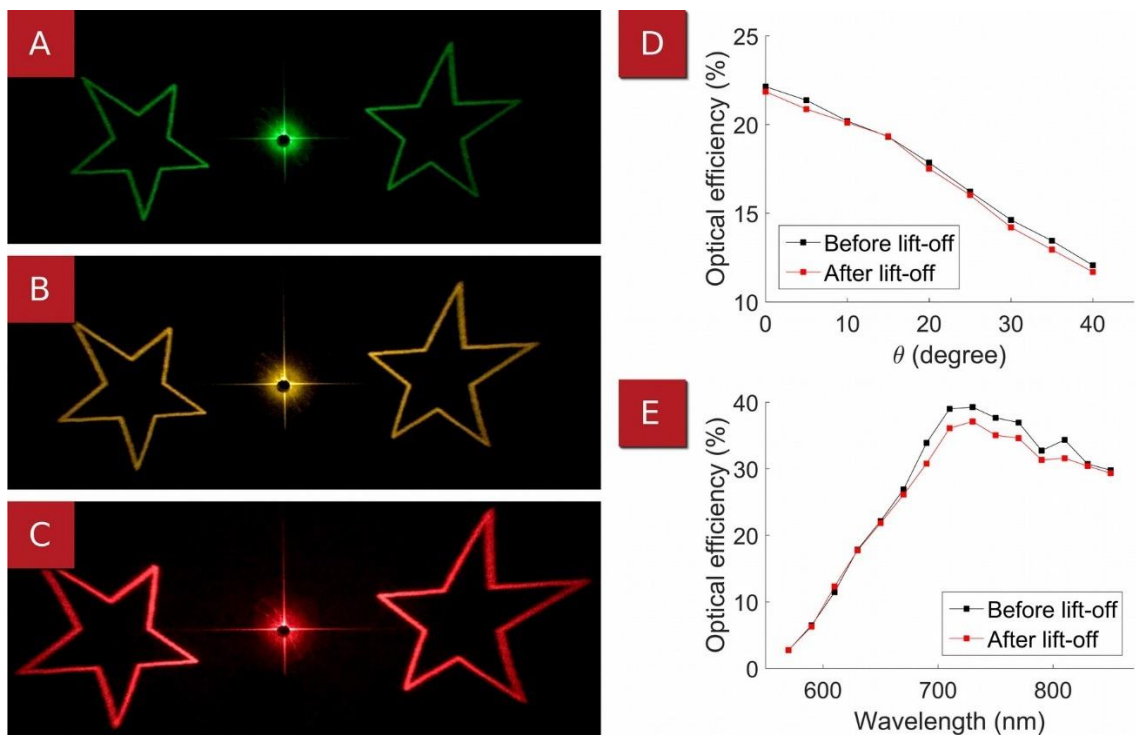

**Fig. 5.** Experimentally obtained images before lift-off for the light beams at **(A)** 570 nm, **(B)** 590 nm, and **(C)** 690 nm. The incident light with right circular polarization impinges on the same MS at normal incidence.

In fig. 6 we placed our MS on a glasses lens with a radius of curvature of 140 mm. This corresponds to a phase change of 1.1 π between the outside and center of the MS. Qualitatively it can be seen in fig. 6B that there is some defocussing effect, and a lower signal to noise ratio than the figs. 4F and 4G. Clearly the effect of curvature can be detrimental to the image quality, but this could also offer opportunities to be investigated in other studies.

A strength of our approach is that we can tailor devices for specific use cases. Here the MS conforms to the general shape of the object it rests on, in this case a glasses lens. The impact of fine surface roughness can be negated by adjusting the thickness of the manipulation layer. A thicker manipulation layer will conform less to the surface roughness and is thus ideal for holography. However, a thinner manipulation layer could be used when a higher degree of conformability to the surface is required. This would be the case for tightly curved surfaces.

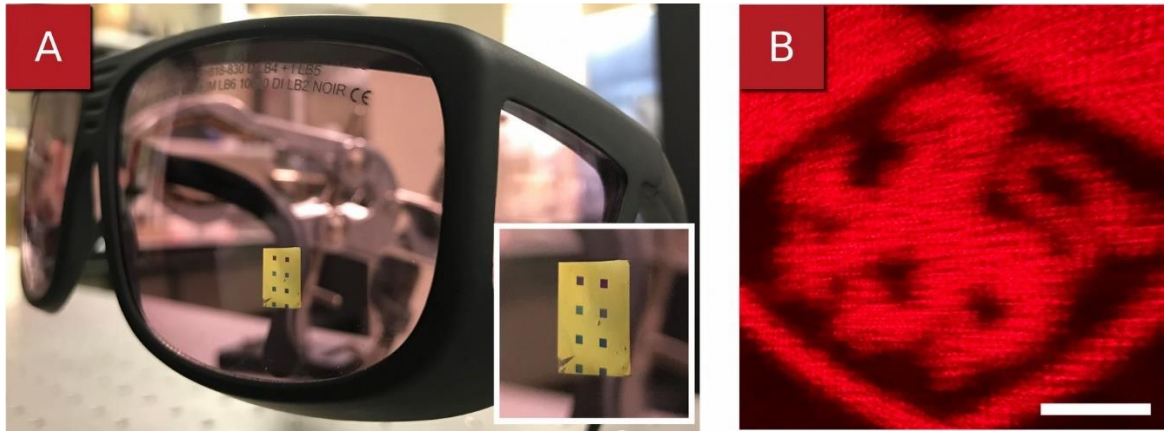

**Fig. 6. (A)** Our MS conformed to a pair of safety glasses. The inset is a close up of the MS. **(B)** An experimentally obtained image with the MS conformed to the glasses for the incident light with a wavelength of 650 nm at normal incidence. The scale bar is 10 mm with a MS to screen spacing of 100 mm.

**Discussion**

We have designed and fabricated a flexible, conformable, holographic MS capable of supporting helicity multiplexed holograms in the visible range. Our MS is realized with gold nanorods positioned point by point to define a Pancharatnam-Berry phase profile. By using a three-layer reflective design on our flexible manipulation layer we can obtain both high efficiency, and conformability to planar and curved surfaces. By then controlling the helicity of the incident light the two high fidelity, broad angle, broadband, images can be interchanged at will.

Such a device is of practical relevance for tunable MS devices with applications including polarization manipulation, beam steering, novel lenses, and holographic displays. Due to the conformability of our device these technologies could be realized on non-planar substrates. Additionally, the flexibility of this device opens the possibility of roll-to-roll printing to vastly decrease the cost, and increase the throughput, of MS production.

**Materials and Methods**

**Fabrication**

We first spin coated, on a silicon carrier, the lift off layer (Omnicoat, from Microchem), at a speed of 1000 rpm for 1 minute, followed by 1 minute of baking at 230 °C.

Then we deposited a thick polymer layer to be used as the MS substrate. For this purpose, we chose SU8, a negative-tone, epoxy based resist from Microchem (MS USA), and available in different formulations. We found that using a blend of SU8 2050 and SU8 2000.5, mixed 1:1 and spun at 5000 rpm gives a thickness of 2.6 µm. This thickness was chosen to guarantee both high flexibility, and robustness of the MS. We exposed this layer to UV light for 5 minutes and completed the cross-linking process by heating the substrate to 100 °C for 5 minutes.

We then deposited, via electron beam evaporation, a 100 nm thick layer of gold, enough to guarantee the efficient reflection of light.

Next we used SU8 to realize the spacing layer, by blending SU8 2000.5 and Cyclopentanone in a 1:3 ratio, and spinning this mixture at 6000 rpm to a thickness of 90 nm. We also exposed this layer to UV light before curing for 2 minutes at 100 °C to cross-link the polymer. We characterized the optical dispersion properties of this film via a standard retrieval method [39], which gave a refractive index of 1.67 at λ=600 nm. The variation in refractive index was less than 1.5% in the wavelength range 570-850 nm.

The top layer of gold for the meta-atoms was 40 nm thick. The meta-atoms were written with a Raith eLINE Plus electron beam device at 30 kV, with a dose of 5 µC/cm2, on a photoresist made of a 1:2 blend of SU8 2000.5 and Cyclopentanone, spun at 5000 rpm and baked for 5 minutes at 90 °C. Ethyl lactate was used to develop the SU8 resist after a post-exposure baking step of 2 minutes at 100 °C.

An Ar based reactive Ion back-etch was then used to etch the gold, leaving nanorods with the dimensions 75 nm×200 nm×40 nm. The parameters of this etch were a DC bias of -325 V, pressure of 0.05 mBar.

The sample was then hard baked at 150 °C for 5 minutes to slightly increase the rigidity of the supporting membrane, to simplify the release step.

To lift-off the MS from the silicon carrier, we used Microposit MF319 to dissolve the Omnicoat layer.

**Supplementary Materials**

section S1. Comparative SEM images of the MS before and after lift-off
section S2. Efficiency correction factor
section S3. Image pre-compensation
fig. S1. SEM image comparison of typical areas of the MS taken before, and after lift-off.
fig. S2. Efficiency correction factor as function of the wavelength.
fig. S3. Holographic target image comparison with and without pre-compensation.
movie S1. Interchange of images with incident laser helicity change.

## Acknowledgments

**Funding:**
We acknowledge support from EPSRC (grants No EP/M508214/1， EP/L017008/1，and EP/M003175/1).
**Author contributions:**
JB carried out the fabrication and performed the experiments. DW and XC designed the hologram and ran simulations. JB wrote the manuscript with contribution from all authors. The work was initiated and directed by ADF and XC.
**Competing interests:**
We declare that they have no competing interests.

## H2: Supplementary Materials

### section S1. Comparative SEM images of the MS before and after lift-off

We took SEM images of the MS before and after lift-off, to assess if the process would affect the quality of the nano-features. Fig. S1A displays an area of the MS before lift-off, and fig. S1B displays a different area of the same MS after lift-off. The quality of the MS is identical in both cases.

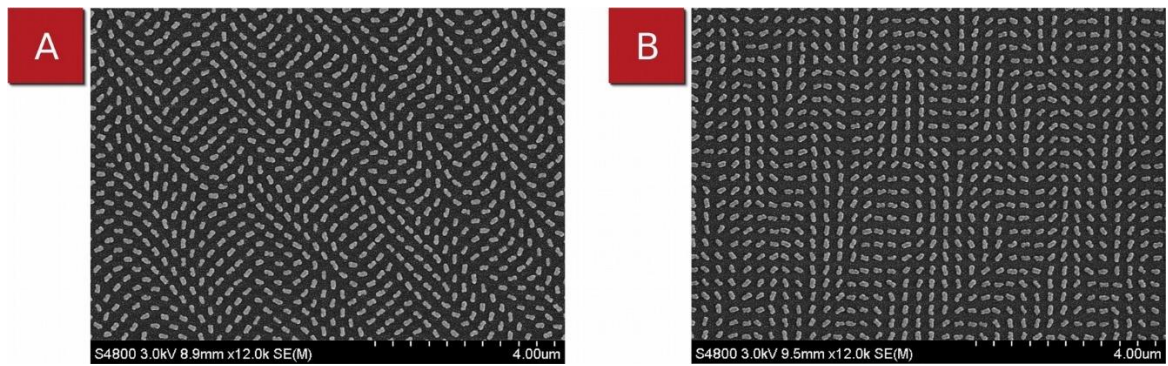

**Fig. S1.** An SEM image of typical areas of the nanorod MS taken **(a)** before lift-off **(b)** after lift-off.

### section S2. Efficiency correction factor

Because the laser beam diameter was larger than the MS, we identified a calibration factor to find the power incident on the MS only. This calibration is a function of the wavelength and is shown in fig. S2. The calibration factor is defined by the ratio of the integral of the beam intensity over the area of the MS and the total intensity. We found this calibration using a Thorlabs BC106N-VIS/M CCD beam profiler to characterize the shape of the beam.

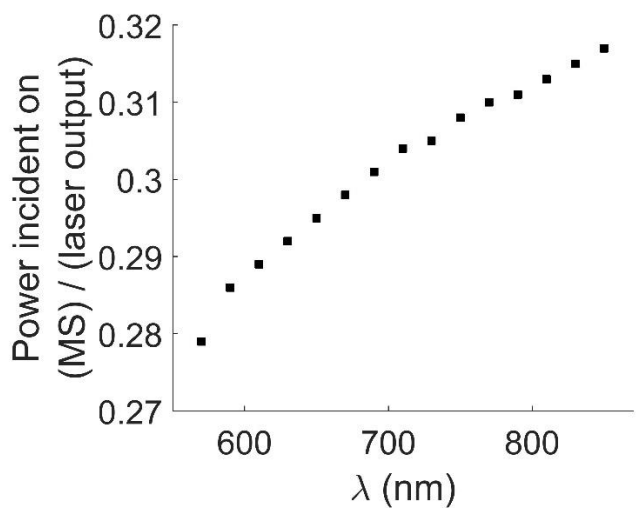

**Fig. S2.** Experimental correction factor as function of the wavelength.

**section S3. Image pre-compensation**
In creating our holographic phase profiles we assume a spherical propagation into the far field. For photographing our images however we use a planar screen. As such, without pre-compensation, the far field image is distorted. To pre-compensate for the distortion, we adjust the target image from fig. S3a to fig. S3b using a spherize filter. The radius of curvature of this filter is chosen to equal that of the far field at the position of the screen, in our case 100 mm. This pre-compensation does effectively lower the resolution at the edges of the image of the target image however. This combined with the non-optimized camera focus contributes to the poorer fidelity of the edges of figs. 3C and 3D.

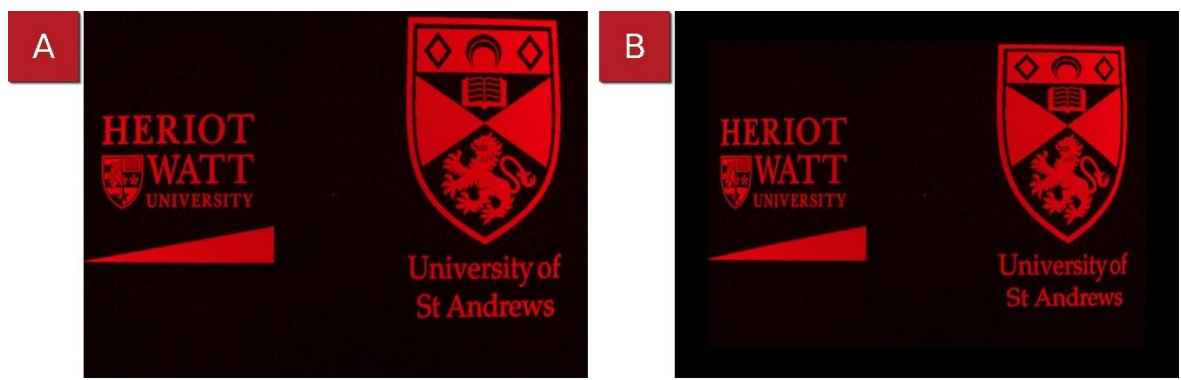

**Fig. S3.** The holographic target image **(a)** without pre-compensation **(b)** with pre-compensation.